# Characterizing the Structure of Topological Insulator Thin Films

Anthony Richardella, Abhinav Kandala, Joon Sue Lee, Nitin Samarth

Department of Physics, The Pennsylvania State University,

University Park PA 16802

We describe the characterization of structural defects that occur during molecular beam epitaxy of topological insulator thin films on commonly used substrates. Twinned domains are ubiquitous but can be reduced by growth on smooth InP (111)A substrates, depending on details of the oxide desorption. Even with a low density of twins, the lattice mismatch between $(Bi,Sb)_2Te_3$ and InP can cause tilts in the film with respect to the substrate. We also briefly discuss transport in simultaneously top and back electrically gated devices using $SrTiO_3$ and the use of capping layers to protect topological insulator films from oxidation and exposure.

Topological insulators (TIs) have attracted much recent attention because of their unusual spin polarized surface states that persist up to room temperature.[1,2] Utilizing these states requires precisely controlling the position of the chemical potential to put it inside the bulk band-gap so that transport can occur solely through the surface states. Since these materials are narrow band-gap semiconductors, unintentional doping is always a concern that can easily move the chemical potential out of the bulk band-gap where the topological transport properties are most pronounced. Further, exposure to atmosphere can also degrade the materials, leading to topologically trivial Rashba-like accumulation regions at the surfaces, thus further complicating the transport properties.[3,4] Therefore, practical uses of the surface state properties of topological insulators require a great deal of attention to minimizing defects during growth and protecting the films after growth and during device fabrication to minimize unintentional changes to the films.

Thin film growth methods, such as molecular beam epitaxy (MBE), are well suited to these materials allowing precise control of the crystal stoichiometry, doping and thickness while also allowing high quality heterostructures to be created to exploit interfacial effects with other materials. $Bi_2Se_3$ and $Bi_2Te_3$ have been the mostly widely studied TI materials by thin film growth. Both have a particularly simple single Dirac cone centered on the Γ point in the Brillouin zone and a rhombohedral (R$\bar{3}$m) crystal structure composed of units of five repeating layers of atoms, such as Se-Bi-Se-Bi-Se, which are called quintuple layers (QLs). Between these QLs, the bonding is of van der Waals type. These weak interfacial bonds make these materials fairly easy to grow using van der Waals epitaxy on many different substrates, even for large lattice mismatches or even different lattice symmetries. Some typical substrates commonly used are GaAs (111), InP (111), Sapphire, Si (111) and $SrTiO_3$ (111), among others.[5–14] These films grow c-axis oriented out of plane, using the typical hexagonal indexing scheme where the (0 0 n) reflections are along the (111) direction if the rhombohedral unit cell was used instead.

The TI heterostructures discussed here were grown by MBE using two chambers connected by an ultrahigh vacuum transfer system. III-V substrates were prepared in the III-V MBE system and then transferred under ultrahigh vacuum to the TI chamber for growth, except in the case were the oxide was desorbed under a Se flux. $Bi_2Se_3$ and $(Bi,Sb)_2Te_3$ films were grown from pure elemental sources of at least 5N purity materials at a base pressure of about 2.5e-10 Torr and monitored by reflection electron high energy diffraction (RHEED). Substrates were indium mounted to the sample holder for growth and typically unmounted in a glove box in argon. Characterization was carried out *ex situ* using a Philips (Panalytical) MRD 4 circle x-ray diffractometer (XRD) in the double axis and triple axis configurations. Transmission electron microscopy was done using an FEI Titan double aberration corrected TEM at 300 keV.

Though TI films will grow on many substrates, the price of the lattice mismatch between them is usually paid by the formation of twin domains in the films. These domains typically consist of 60° rotations in plane around the c-axis. Given the rhombohedral symmetry, the surface likes to form triangular structures and twinned domains are readily observed, using atomic force microscopy (AFM) for instance, by the points of the triangles pointing in opposite directions (Fig. 1(a)). Of these substrates, InP (111) is particularly well lattice matched to $Bi_2Se_3$ and significant reductions of twin domains have been observed by several groups using either InP (111)A or (111)B substrates, where one orientation with respect to the substrate lattice is believed to be stabilized by nucleation of the film at step edges of the substrate.[6,8] Consistent with this, we observe a large suppression of one orientation when growing $Bi_2Se_3$ on InP (111)A substrates were the oxide was desorbed prior to growth at ~450 C (measured by band edge thermometry) under a flux of arsenic to help prevent phosphorus out-diffusion (Fig 1(b)). After desorbing under As the InP (111)A RHEED pattern shows a clear 2x2 reconstruction which changes to a 1x1 pattern as soon as the surface is exposed to Se. The $Bi_2Se_3$ film was grown at a temperature above where RHEED oscillations can no longer be observed (substrate temperature of 390 C, pyrometer temperature of 330 C), in the step-flow growth mode.[13] Interestingly, when we desorb the substrate under a flux of Se, instead of As, much of the suppression of one of the orientations is lost and in fact the opposite orientation is now preferred (Fig 1(c)). Note that the InP (113) plane occurs at the same φ azimuthal angle as the (002) plane referred to in Ref 8. We also observe that a partial suppression of twinning can also occur on sapphire substrates, despite to larger lattice mismatch, as shown in Fig 1(d) for an 8 QL thick $Bi_2Se_3$ film. The $Bi_2Se_3$ rocking curve on sapphire shows a sharp distribution and a broad one, indicating a distinct part of the film is more ordered than the rest, similar to earlier observations.[15]

Due to the tendency of $Bi_2Se_3$ to form Se vacancies, it is easier to make insulating films of $(Bi_xSb_{1-x})_2Te_3$. Like $Bi_2Se_3$, $Bi_2Te_3$ tends to be n-type while $Sb_2Te_3$ tends to be p-type and combining them to create the alloy makes a naturally compensated material. The ability to place the chemical potential at the Dirac point, along with out-of plane ferromagnetism being possible by Cr or V doping, make this one of the most promising compounds for exotic effects such as the quantum anomalous Hall effect.[16–19] $(Bi,Sb)_2Te_3$ shares the same twinning problems as $Bi_2Se_3$ but similarly, growth on InP (111)A strongly favors one orientation (Fig 2(a)). The lattice mismatch is larger however and we have observed on thicker films of ~100 QL that the rocking curve can split into two distinct peaks. Fig 2(b) shows the rocking curve as a function of azimuthal angle for a second Cr doped film and Fig 2(c) shows the corresponding reciprocal space map of the same (0 0 6) reflection. All the (0 0 n) reflections shift to slightly higher angle, consistent with a decrease in the lattice c-axis lattice constant due to the substitutional incorporation of Cr.[20] The composition was determined to be $Cr_{0.188}(Bi_{0.4},Sb_{0.6})_{1.812}Te_3$ by secondary ion mass spectrometry (SIMS). The broad diffuse horizontal background in the reciprocal space map indicates mosaic disorder, while the two peaks with Kiessig thickness fringes in the vertical direction indicate two regions of the film that are tilted slightly away from the substrate (1 1 1) normal in opposite directions. Taken together, it appears the film is able to accommodate the mismatch by buckling, forming a corrugation predominantly along one direction. Tilting of Te-based films has also been observed on vicinal substrates, likely associated with the high step density.[21] A high angle annular dark field (HAADF) scanning transmission electron microscope (STEM) image of the sample is displayed in Fig 2(d) showing the lattice is generally well ordered. One dislocation, of the type responsible for the mosaic texture, can be seen running from the substrate to the top of the film marked by a red arrow. Elemental mapping using energy dispersive spectroscopy (EDS) was not able to resolve any evidence of elemental clustering, although it is possible that some indium from the substrate does diffuse into the $(Bi,Sb)_2Te_3$ film near the interface. It has been shown that strain in the crystal near the low angle grain boundaries modifies the Dirac states and can even open a gap under compressive strain.[22] Therefore the further reduction of the mosaic spread of such films is highly desirable.

It is desirable to electrically gate devices made from such films to move the chemical potential across the Dirac point. Typically top gates are defined using lithography. However, it is well known such processes can degrade the quality of the crystals. Another standard approach to gating is to grow on

SrTiO$_3$ (STO) (111) substrates.[14,16] STO substrates require annealing in oxygen at high temperatures to form an atomically ordered surface consisting of single atomic steps. We typically clean the as received substrates in acetone, isopropanol and DI water before annealing them in an oxygen rich environment at ~900 C to 950 C for 2.5 hours to accomplish this. We then check the annealed surfaces with AFM before using them. Substrates with a high density of steps and terraces less than 10 nm apart tend not to work well for growth of (Bi,Sb)$_2$Te$_3$. Fig. 3(a) displays an AFM image of a nominally ~1.5QL thick film showing that the film nucleates as islands that slowly merge together, as they have in some regions. It is clear there are voids even in regions where the islands have merged. In fact, a histogram of the height profile shows that bare substrate and 2, 3 and 4 QL thick film islands are identifiable but single QL thick regions are almost absent, indicating that the film does not like the polar STO surface.[23] Two step growth using a low temperature deposited seed layer does not appear to work either for this reason. Nevertheless, once the islands merge after growing a thicker layer a decent film can be formed. The addition of Cr doping also appears to help the film adhere due to the high sticking coefficient of Cr. As seen in the STEM image in Fig 3(b) an amorphous layer exists at the interface followed by a well ordered film. As a consequence of the nucleation of the film, however, there are blobs that are seen on the surface (Fig 3(c)). These defects are perhaps misoriented (Bi,Sb)$_2$Te$_3$ grains based on the fact that they are crystalline and EDS cannot distinguish the composition from the rest of the film (Fig 3(d)).

Such films are easily mechanically scratched and patterned into Hall bars that can be back gated without any lithography required.[16] This creates a stronger electric field at one surface of the film than the other creating a gradient in the chemical potential across the film. Gating both the top and back surface of the film independently therefore is a potentially promising approach for effects such as the quantum anomalous Hall effect, where E$_F$ needs to be in the magnetic surface state gap of both surfaces simultaneously. In this direction, we fabricated a dual gated thin film device of Cr-(Bi,Sb)$_2$Te$_3$ grown on STO. The film was first patterned into a Hall bar by mechanical scratching, followed by atomic layer deposition of an Al$_2$O$_3$ dielectric layer for a Ti/Au top gate that was subsequently defined by photolithography. The film in this case was inherently p-type. With the back gate of 80 V, the peak in R$_{xx}$, indicating gating across the Dirac point, occurred with a top gate near zero. As expected, with a less positive back gate voltage, a more positive top gate voltage was necessary to traverse the Dirac point, as shown in the resistance map of Figure 4(a). The lithography process however degrades the size of the anomalous Hall effect measured making it significantly smaller than the quantized limit.[24] Nevertheless, we observe a systematic increase in the coercive field as the film is gated into the valence band consistent with observations that the magnetism may be carrier mediated in the bulk bands (Fig 4(b,c)).[25,26]

Finally, we briefly mention our observations with the passivation of thin films using capping layers. Capping TI films to protect them from exposure to ambient conditions has been widely employed, with Se, Te and Al caps probably the most commonly used materials. A thin Al cap oxidizes immediately on exposure to air leaving an inert layer on the surface.[27] Efforts to see if such an Al cap can protect during the fabrication of a dual gated devices are ongoing. Se caps are more useful when the cap needs to be removed to expose the clean surface again for ex situ measurements or for fabrication. Se is only useful for Bi$_2$Se$_3$ films, however, as Se will readily diffuse into Te based films, changing their properties.[15] In our experience, Te caps cannot be removed cleanly after they have been exposed to air. Typically, Se is deposited after the sample has cooled to room temperature as an amorphous layer. Alternately, if the flux of Se is left on as the sample cools to room temperature, we observed by XRD that rods of crystalline hexagonal Se form instead with their c-axis lying in-plane. Such a capped sample and the surface after desorbing are shown in Fig 5(a). Compared to an amorphous Se cap, the crystalline one is more robust and stands up better to heating in air, whereas (Fig. 5(b)) heating a sample with a 50 to 100 nm thick amorphous cap near to 100C quickly results in the cap balling up and exposing parts of the underlying film. The decapping procedure in vacuum is identical regardless of whether the Se cap is amorphous or crystalline, as indeed, it probably becomes crystalline as it is heated. If the samples are stored in an inert environment, we do not find that a pre-sputter cleaning of the Se cap is necessary to recover a clean surface.[28] Typically we heat the sample slowly to ~200 C and about 10 minutes after the film RHEED pattern is recovered the surface is completely clean.

In conclusion, we have characterized some of the structural defects that occur in TI films grown on commonly used substrates. The lattice mismatch with the substrate and the nature of the states at the interface can have a significant effect on the nucleation of the film and on the formation of twinned domains within it. Despite the complication of the poor initial film layer using STO substrates, it is an extremely flexible material for gated studies of transport in TI materials. More complicated devices that require lithographic patterning will clearly benefit from improved ways of capping these materials that prevent the unintentional degradation of their material properties, particularly in magnetically doped samples.

This work was supported by grants from ONR (N00014-12-1-0117), ARO MURI (W911NF-12-1-0461), DARPA (N66001-11-1-4110) and by the C-SPIN center, one of six STARnet program research centers. We also acknowledge use of the NSF National Nanofabrication Users Network Facility at Penn State.

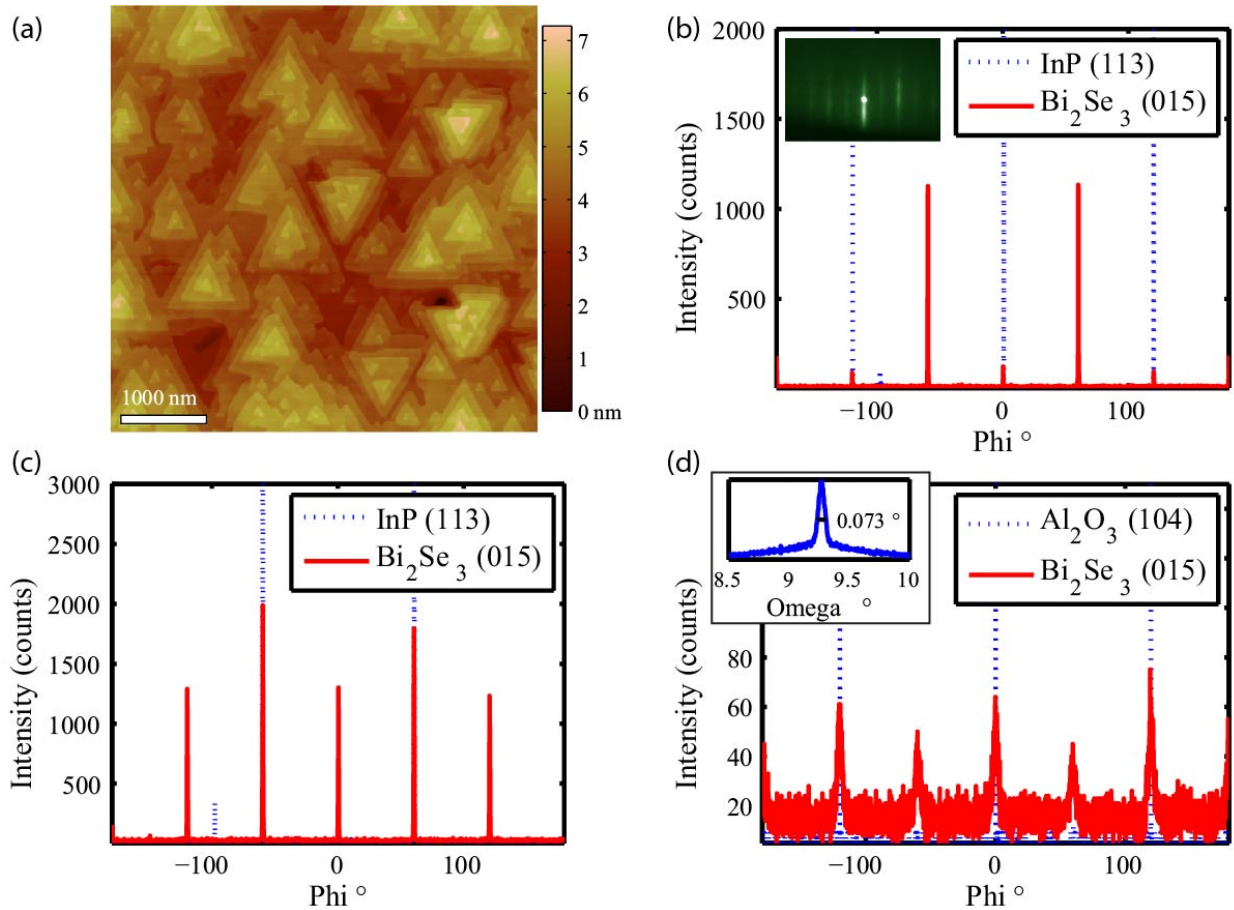

Figure 1 : (a) AFM image of Bi$_2$Se$_3$ on InP (111)A where the oxide was desorbed under a flux of As. Occational regions with triangular structures pointing down are twinned domains. (b) XRD phi scan of sample in (a) showing the film is predominantly single crystal. Inset is a RHEED image of the InP substrate desorbed under As flux before growth showing a 2 x 2 reconstruction. (c) Phi scan of Bi$_2$Se$_3$ on InP (111)A desorbed under a Se flux. The reduction is twins is not as effective and the other orientation with respect to the substrate is preferred. (d) Phi scan of an 8 QL thick Bi$_2$Se$_3$ film on sapphire showing twins are also reduced on this substrate. Double axis rocking curve of the (0 0 6) reflection in inset shows evidence two distributions, one significantly broader and more disordered than the other.

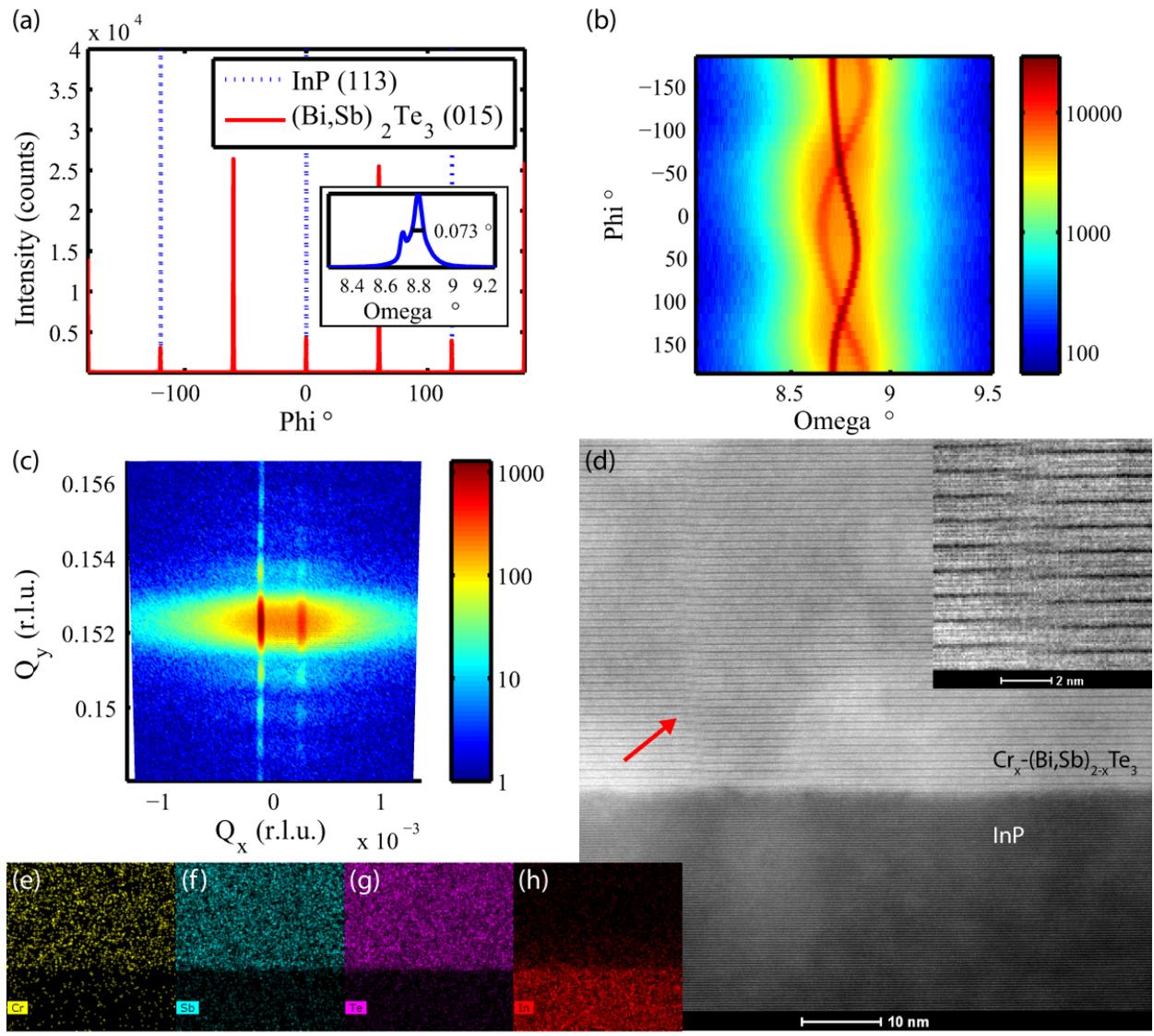

Figure 2: (a) XRD phi scan of a 100 QL thick $(Bi,Sb)_2Te_3$ film on InP (111)A desorbed under As. Inset shows a splitting of the double axis rocking curve of the (0 0 6) reflection, which is sometimes observed for thicker films such as this. (b) Phi-omega scan for an 83 QL Cr-$(Bi,Sb)_2Te_3$ film showing the rocking curve splitting is due to the twinned domains tilting on opposite directions to accommodate the lattice mismatch with the substrate. (c) Corresponding triple axis reciprocal space map for sample in (b) showing the two peaks and a wide diffuse mosaic spread. (d) HAADF STEM image of the same sample showing the sample is well ordered locally. A dislocation running from the substrate to the top of the film is zoomed in on in the inset from the area indicated by the red arrow. (e-h) EDS elemental maps of the area in (d) for Cr, Sb, Te and In respectively. No clustering near dislocations was observed.

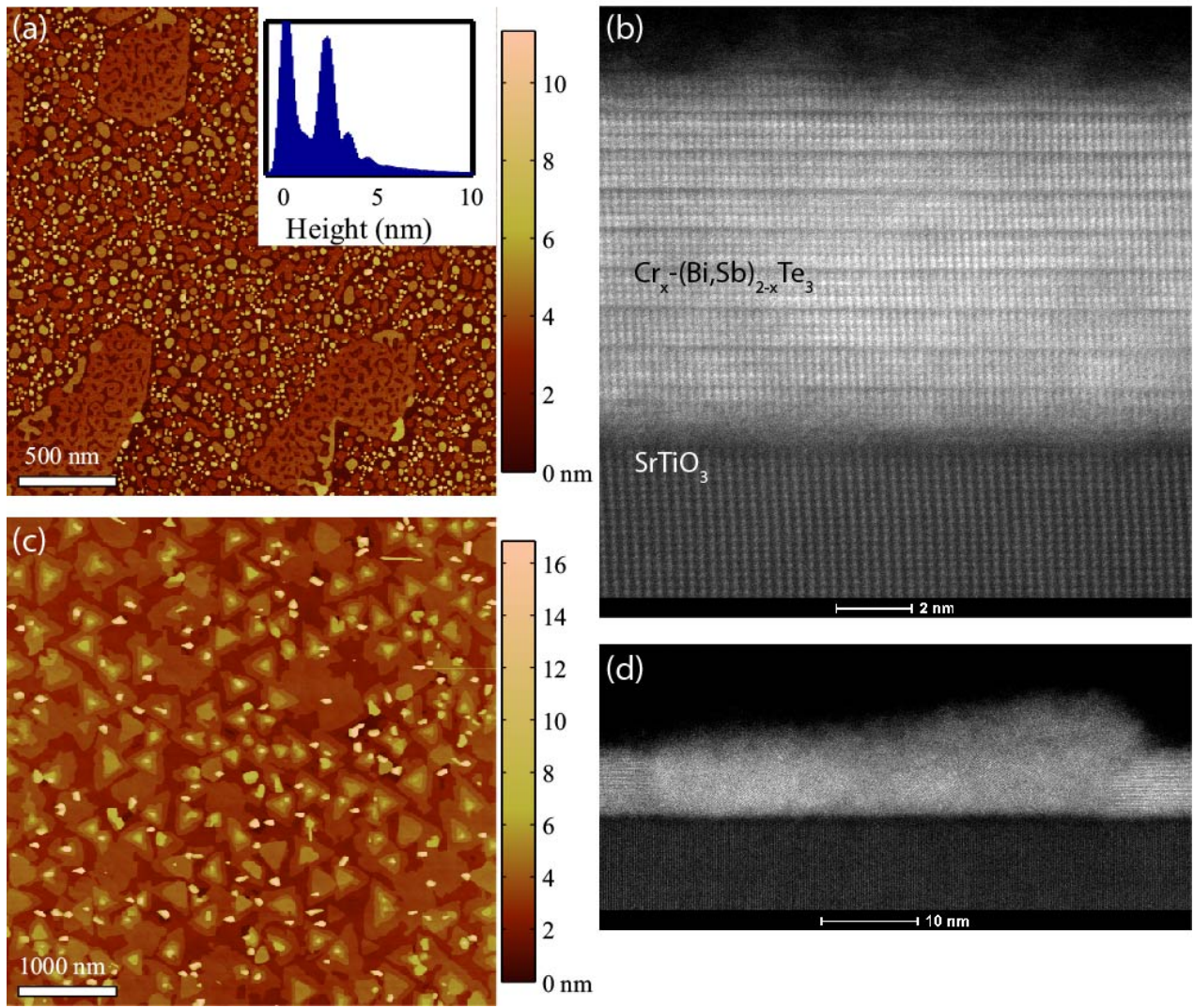

*Figure 3: (a) AFM image of an ~1.5 QL thick (Bi,Sb)$_2$Te$_3$ film on an SrTiO$_3$ substrate showing the film nucleates as islands that merge together. Histogram in inset shows that bare STO and 2 QL thick films regions are most common. (b) HAADF STEM of a Cr-(Bi,Sb)$_2$Te$_3$ film showing a well ordered film with an amorphous interfacial region at the interface. (c) AFM of an 8 QL Cr-(Bi,Sb)$_2$Te$_3$ film on STO. A number of blobs can be seen on the surface that are up to 20 nm tall. (d) Feature likely associated with the blobs seen in AFM. Appears crystalline and EDS confirmed a similar composition as the rest of the film leading us to speculate it may be a misoriented Cr-(Bi,Sb)$_2$Te$_3$ grain.*

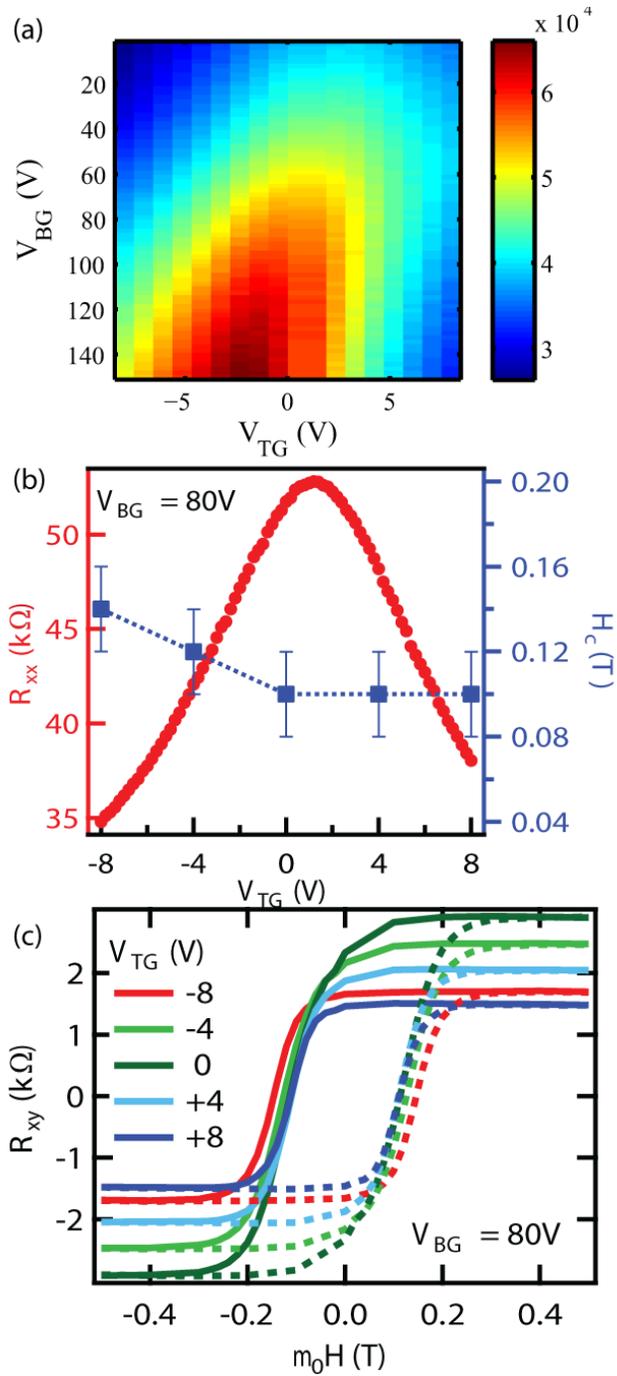

Figure 4: (a) $R_{xx}$ resistance as a function of top and back gate voltages of a Cr-$(Bi,Sb)_2Te_3$ film on $SrTiO_3$. (b) $R_{xx}$ vs. top gate (red) at a fixed back gate of 80 V showing the maximum in resistance at the Dirac point. The coercive field vs. gate measured from $R_{xy}$ is plotted (blue) against the right axis. (c) $R_{xy}$ hysteresis loops as a function of top gate voltage at fixed back gate.

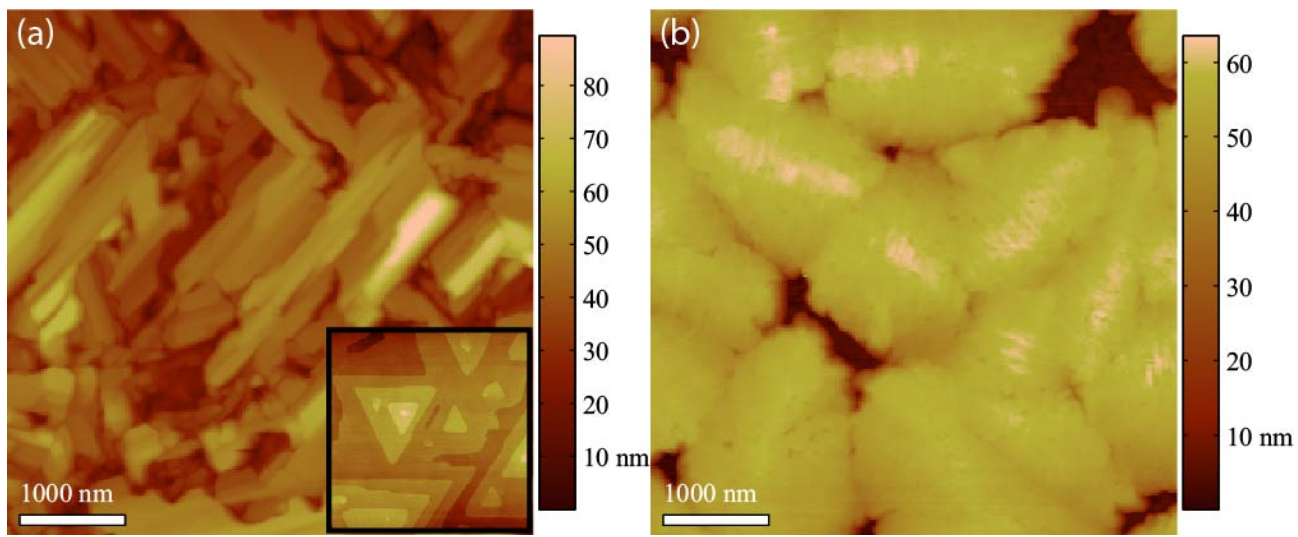

*Figure 5: (a) AFM image of a crystalline Se capping layer on $Bi_2Se_3$. Inset shows a clean surface after thermally desorbing the capping layer in vacuum. (b) AFM image of an amorphous Se cap that was briefly heated in air. In contrast to the crystalline cap, the Se clusters up leaving voids behind that expose the $Bi_2Se_3$ film.*